\documentclass[twocolumn,showpacs,preprintnumbers,amsmath,amssymb, epsfig, prb]{revtex4}
\usepackage{graphicx}
\usepackage{dcolumn}
\usepackage{bm}

\begin{document}

\title{Enhancement of the Curie temperature in GaMnAs/InGaMnAs superlattices}
\author{A. Koeder,$^{a)}$\footnote[0]{$^{a)}$Electronic-mail: achim.koeder@physik.uni-ulm.de} W. Limmer, S. Frank, W. Schoch, V. Avrutin, R.Sauer, and A. Waag}
\affiliation{Abteilung Halbleiterphysik, Universit\"{a}t Ulm, D-89069 Ulm, Germany}
\author{K. Zuern, and P. Ziemann}
\affiliation{Abteilung Festk\"{o}rperphysik, Universit\"{a}t Ulm, D-89069 Ulm, Germany}


\begin{abstract}
We report on an enhancement of the Curie temperature in GaMnAs/InGaMnAs superlattices
grown by low-temperature molecular beam epitaxy, which is due to thin InGaMnAs or InGaAs
films embedded into the GaMnAs layers. The pronounced increase of the Curie temperature
is strongly correlated to the In concentration in the embedded layers. Curie temperatures
up to 110 K are observed in such structures compared to 60 K in GaMnAs single layers
grown under the same conditions. A further increase in T$_C$ up to 130 K can be achieved
using post-growth annealing at temperatures near the growth temperature. Pronounced
thickness fringes in the high resolution X-ray diffraction spectra indicate good
crystalline quality and sharp interfaces in the structures.
\end{abstract}


\maketitle

Developing ferromagnetic semiconductors with Curie temperatures near room temperature or
beyond is of major interest for future spintronic devices. One material under intense
investigation in recent years is the diluted magnetic semiconductor Ga$_{1-x}$Mn$_x$As
which can be made ferromagnetic at present with Curie temperatures up to 110
K\cite{Matsukuratransport} in as-grown samples and up to 150 K\cite{Kuannealing} in
annealed samples. GaMnAs is fabricated by low-temperature (LT) molecular beam epitaxy
(MBE) with growth temperatures of around 250$^{\circ}$C, and therefore, a lot of defects
are incorporated into the lattice. Different defect species like As antisites, As$_{Ga}$,
or As interstitials, As$_I$, as well as Mn interstitials, Mn$_I$, are discussed in the
literature,\cite{Potashnik}$^-$\cite{Blinowski_interstitials} having one aspect in
common: They are all acting as donors and are therefore compensating the Mn acceptors in
GaMnAs. Hence, the defect structure influences strongly the electrical and, consequently,
the magnetic properties of this material system, since the ferromagnetic coupling in
GaMnAs is suggested to be due to an indirect exchange interaction between the magnetic
moments of Mn$^{++}$-ions mediated by holes.\cite{DietlScience287zenermodell} So, one
possible way to increase the Curie temperature T$_C$ in GaMnAs is to increase the hole
concentration, e.g., by reducing the defect density in the material by post-growth
annealing at temperatures near or even below the growth
temperature.\cite{Kuannealing}$^,$\cite{Foxon} Another possibility to change the magnetic
properties is to control the defect structure of the material even during
growth.\cite{Tanaka_epitaxial}$^-$\cite{Campion_epitaxial}


Here, we report on the increase of T$_C$ of GaMnAs due to strained
(In$_y$Ga$_{1-y}$)$_{1-x}$Mn$_x$As or In$_y$Ga$_{1-y}$As layers embedded into the
magnetically active GaMnAs layers.

All samples studied were grown on epiready semi-insulating GaAs(001)  substrates in a
Riber 32 MBE system. To provide As$_4$, an As cracker cell was used in the non-cracking
mode. To supply Ga and In fluxes, conventional Knudsen cells were used, whereas a Hot-Lip
cell was used for Mn. The growth procedure of the GaMnAs/InGaMnAs superlattices was as
follows: First, a 100 nm thick GaAs buffer layer was grown at a substrate temperature of
T$_S$ = 585$^{\circ}$C (standard GaAs growth conditions). Then the sample was cooled down
during a growth break to the growth temperature of T$_S$ = 230$^{\circ}$C. The growth was
continued by a 20 nm thick Ga$_{1-x}$Mn$_x$As layer followed by a 3-6 nm thick
(In$_y$Ga$_{1-y}$)$_{1-x}$Mn$_x$As layer. This LT procedure was repeated 15 times.
Finally, the structure was capped by an additional 20 nm thick Ga$_{1-x}$Mn$_x$As layer.
The In content in the InGaMnAs films was varied from sample to sample between zero and
50\%. Mn-flux was kept constant during the whole LT-growth and corresponded to a Mn
content of $x$ = 0.058 in the Ga$_{1-x}$Mn$_x$As layers. In the
(In$_y$Ga$_{1-y}$)$_{1-x}$Mn$_x$As layers it was somewhat less, depending on the In
content in the sample. The same growth procedure was used to grow the
GaMnAs/In$_y$Ga$_{1-y}$As superlattices except the missing Mn in the embedded layers. The
growth was monitored in situ by reflection high energy electron diffraction (RHEED). The
RHEED pattern showed (1 $\times$ 2) reconstruction during the growth of the whole LT part
of the structure and no evidence of a second phase on the surface (i.e. MnAs-clusters)
was observed. Details of the growth of GaMnAs single layers is also described
elsewhere.\cite{Koeder_gradient} Annealing experiments were performed in air using a
LINKAM THMS 600 heating chamber. The annealing temperature was chosen to be
250$^{\circ}$C, and the annealing time was 30 min. Details of the annealing experiments
are described elsewhere.\cite{wolfgang_annealing}

\begin{figure}
\includegraphics[width=0.5\textwidth]{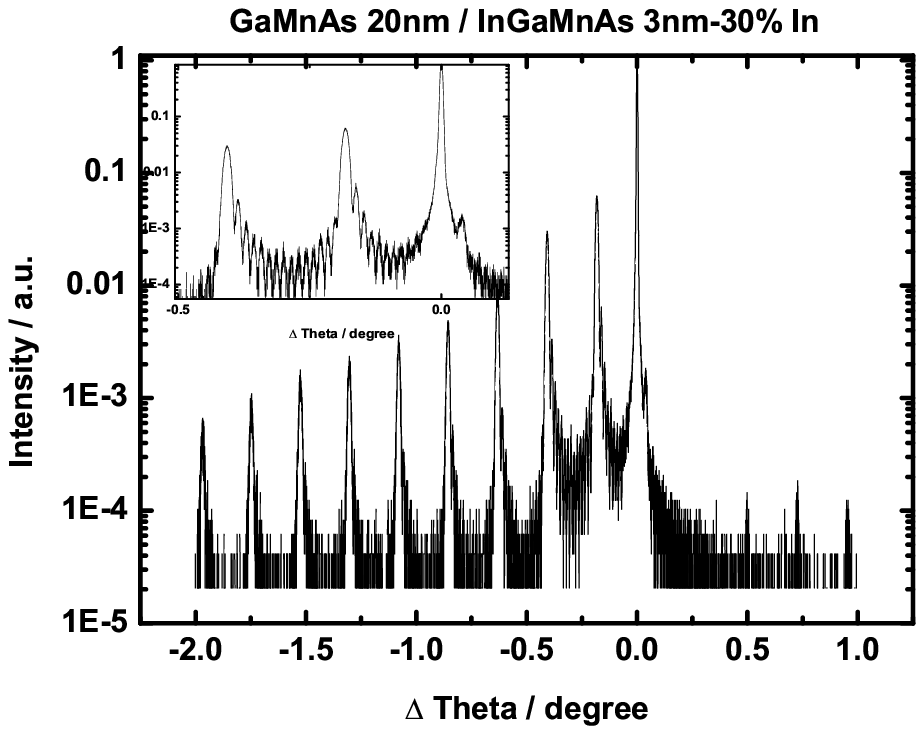}
\caption{\label{HRXRD_SL}$\theta$-$2\theta$-scan of a
Ga$_{0.942}$Mn$_{0.058}$As/(In$_{0.3}$Ga$_{0.7}$)$_{0.96}$Mn$_{0.04}$As} superlattice.
The inset shows pronounced thickness fringes between the zero order peak and the second
order peak.
\end{figure}

For all superlattices we performed high resolution X-ray diffraction (HRXRD)
measurements. As seen for example in Fig.~\ref{HRXRD_SL}, the $\theta$-$2\theta$-scans
indicate good crystalline quality and sharp interfaces between the GaMnAs and the
InGaMnAs or InGaAs layers. Up to ten superlattice peaks can clearly be resolved, and even
between the peaks pronounced thickness fringes are observed (see inset
Fig.~\ref{HRXRD_SL}). From the superlattice fringes, the period of the superlattices can
be caluclated. In the particular example, the 23.6 nm derived from HRXRD are in good
agreement to what we expect from the growth rates. Finally, the sharp symmetrical peaks
indicate a fully strained structure.

\begin{figure}
\includegraphics[width=0.5\textwidth]{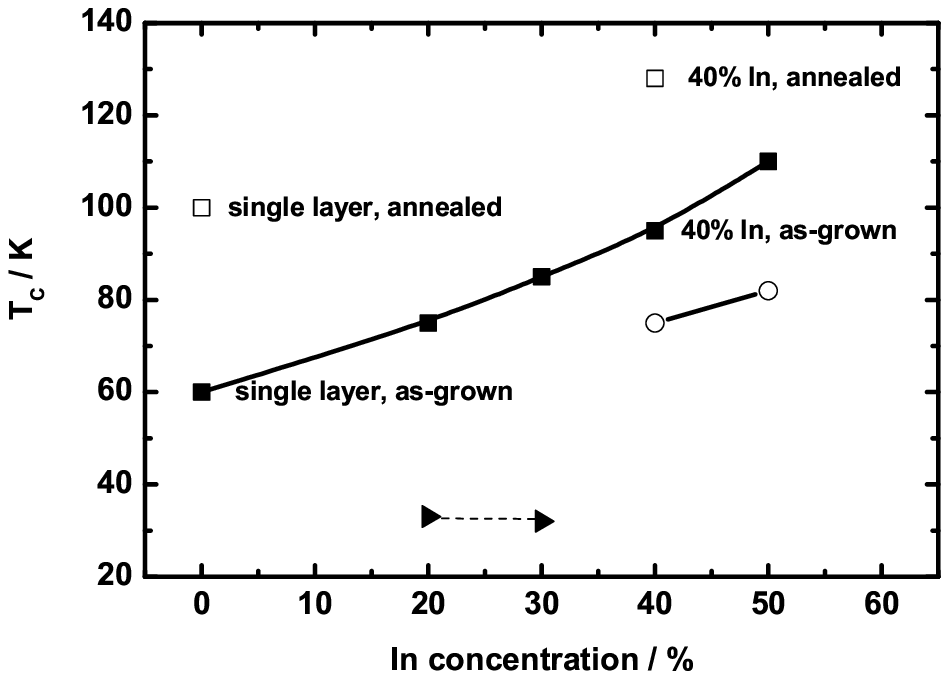}
\caption{\label{tc_vs_in_sl}} Curie temperature as a function of In-content in
GaMnAs/InGaMnAs superlattices. Filled squares: as-grown GaMnAs/InGaMnAs, open squares:
annealed GaMnAs/InGaMnAs, open circles: as-grown GaMnAs/InGaAs, filled triangles:
as-grown LT-GaAs/InGaMnAs. The lines are to guide the eye.
\end{figure}

A series of samples containing GaMnAs/InGaMnAs superlattices were grown with varying In
content from 0\% In (single layer) up to 50\% In. To extract the Curie temperature of the
samples superconducting quantum interference device (SQUID) magnetization measurements
were performed as a function of temperature (accuracy $\pm$ 5 K) in a small in-plane
applied magnetic field (50 Oe $\|$ (110)). As can be seen in Fig~\ref{tc_vs_in_sl}, T$_C$
is continously increasing with increasing In content (filled squares) from 60 K in the
GaMnAs single layer up to 110 K  in the as-grown structure containing 50\% In.

In order to further analyze this effect, several reference samples were grown under the
same growth conditions: First, two LT-GaAs/InGaMnAs superlattices containing 20 and 30\%
In, in order to clarify whether the higher Curie temperature stems from the GaMnAs or the
InGaMnAs layers. These two samples, in Fig.~\ref{tc_vs_in_sl} marked by filled triangles,
show much lower Curie temperatures of around 30K. Therefore, the InGaMnAs films alone
cannot be responsible for the high Curie temperatures in the superlattices. Second, two
GaMnAs/InGaAs superlattices containing 40 and 50\% In (open circles in
Fig.~\ref{tc_vs_in_sl}). These two samples show enhanced Curie temperatures of 70 and 80
K, respectively, even though there is no Mn in the InGaAs.  The reason for the
differences in T$_C$, measured for the superlattices with and without Mn in the InGaMnAs
layer, is not yet known.

Obviously, the high Curie temperatures of the superlattices are due to an enhancement of
T$_C$ in the GaMnAs layers by incorporating strained In-containing layers.

A further increase in T$_C$ was observed upon annealing the samples 30 min at
250$^{\circ}$C, which is near the growth temperature. Finally, a Curie temperature of 100
K was observed for the annealed GaMnAs single layer and 130 K for the annealed
superlattice containing 40\% In (open squares in Fig.~\ref{tc_vs_in_sl}).

\begin{figure}
\includegraphics[width=0.4\textwidth]{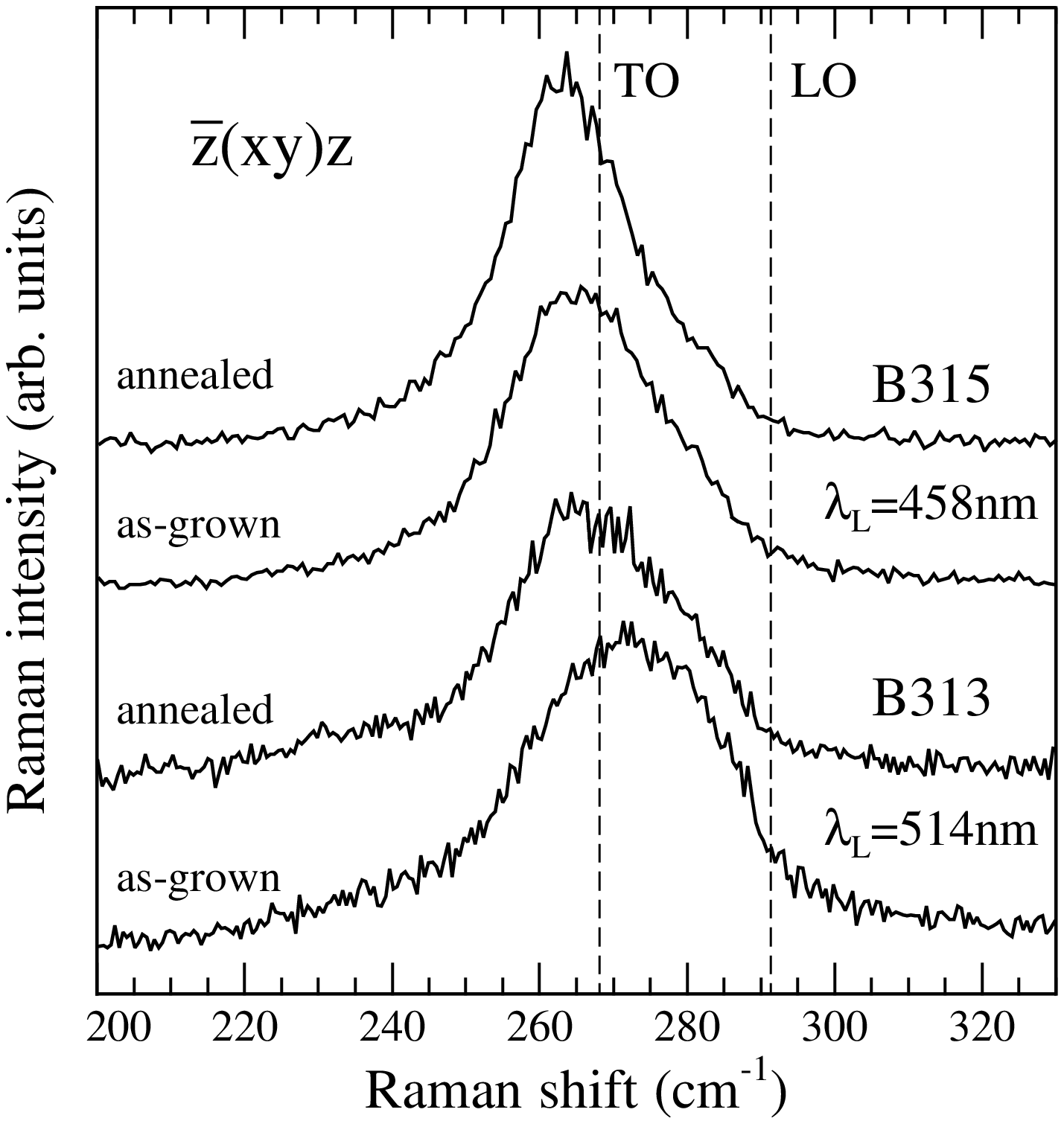}
\caption{\label{raman_sl}} Raman spectra recorded from a GaMnAs single layer (B313)and a
GaMnAs/InGaMnAs superlattice with 40\% In (B315) before and after annealing.
\end{figure}

The hole concentrations of heavily doped GaMnAs layers can be estimated by a careful
inspection of the coupled plasmon-LO-phonon modes, observed in Raman spectroscopy.
Details of this method are described elsewhere.\cite{LimmerRaman} In Fig.~\ref{raman_sl},
the Raman spectra recorded from the GaMnAs single layer (sample B313) and from the
GaMnAs/InGaMnAs superlattice with 40\% In (sample B315) before and after annealing are
depicted for comparison. All four spectra exhibit a broad Raman line, located near the
frequency of the GaAs TO phonon, which arises from the coupled mode. With increasing hole
concentration this line is known to shift from the frequency of the LO phonon to that of
the TO phonon. Therefore, the Raman spectra reveal two features: First, an increase of
the hole concentration upon annealing in both samples, and second, a significant
enhancement of the hole concentration in the GaMnAs region of the superlattices compared
to the GaMnAs single layer.

\begin{figure}
\includegraphics[width=0.5\textwidth]{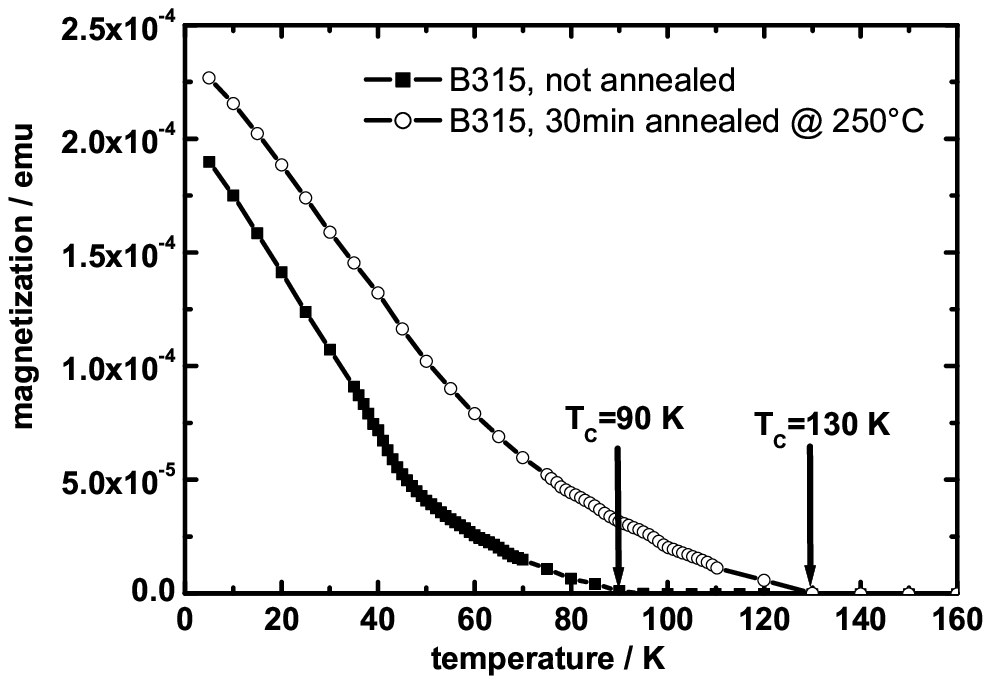}
\caption{\label{squid_sl}} SQUID magnetization curves of a superlattice containing 40\%
In in the embedded layer before and after annealing at 250$^{\circ}$C.
\end{figure}

This agrees very well with the observed increase in T$_C$. Figure~\ref{squid_sl} shows
the SQUID magnetization curves of the superlattice in Fig.~\ref{raman_sl} (sample B315)
before and after annealing. The increase in Curie temperature and also in saturation
magnetization due to annealing can clearly be seen.

At present, the underlying mechanism of the drastic enhancement of T$_C$ in the
GaMnAs/InGaMnAs heterostrucures is still unclear.  However, we suggest that the
incorporation of strained InGaMnAs or InGaAs films into GaMnAs leads to a reduction of
the density of compensating defects, resulting in a higher hole density, and thus, in a
higher Curie temperature than in GaMnAs single layers. This assumption is in agreement
with the Raman data which revealed an increase of the hole density in the GaMnAs barrier
layers of the superlattices compared to the GaMnAs single layer. From recent
publications\cite{Yuinterstitials}$^,$\cite{Blinowski_interstitials} it is known, that
the defects, which are most likely responsible for carrier compensation in GaMnAs, are
the Mn atoms on interstitial sites, and LT-annealing can reduce their density. Edmonds et
al.\cite{Edmonds_annealing_interstitials} demonstrated that out-diffusion of Mn$_I$
towards the surface takes place during annealing of GaMnAs single layers. In
GaMnAs/InGaMnAs superlattices, the In-containing layers and/or interfaces can act as
sinks for point defects like Mn$_I$ and As$_I$ during the growth. Most likely diffusion
of Mn$_I$ towards the heterointerfaces is responsible for the increase in carrier density
and, consequently, T$_C$.  Since in our superlattices a further enhancement of T$_C$ can
be achieved by post-growth annealing, too, a considerable amount of Mn$_I$ ions seems to
be still present in the as-grown superlattices. In contrast to GaAs/GaMnAs interfaces,
which have been shown to prevent removal of Mn$_I$ during annealing,\cite{Kuannealing}
the embedded InGaMnAs layers obviously do not have any adverse effect on the annealing
process.

In summary we have shown that the Curie temperature of epitaxially grown GaMnAs layers
can be increased by embedding strained InGaMnAs or InGaAs layers into the magnetically
active GaMnAs. We suggest that this is due to the influence of strain on the diffusion of
defects like Mn and As interstitials into the In containing embedded layers, or to the
reduction of their concentration during growth as compared to GaMnAs single films. The
Curie temperature can be controlled by varying the In content in the embedded layers.

The authors acknowledge the financial support by the Deutsche Forschungsgemeinschaft, DFG
Wa 860/4-1.

\end{document}